\documentclass[prb,showpacs,amsmath,amssymb,superscriptaddress,twocolumn,
floatfix]{revtex4}

\usepackage[english]{babel}
\usepackage{graphicx}
\usepackage{times}

\begin{document}

\title{Energy resolution and discretization artefacts in the numerical
renormalization group}

\author{Rok \v{Z}itko}
\affiliation{Institute for Theoretical Physics, University of G\"ottingen,
Friedrich-Hund-Platz 1, D-37077 G\"ottingen, Germany}
\affiliation{J. Stefan Institute, Jamova 39, SI-1000 Ljubljana, Slovenia}

\author{Thomas Pruschke}
\affiliation{Institute for Theoretical Physics, University of G\"ottingen,
Friedrich-Hund-Platz 1, D-37077 G\"ottingen, Germany}

\date{\today}


\begin{abstract}
We study the limits of the energy resolution that can be achieved in the
calculations of spectral functions of quantum impurity models using the
numerical renormalization group (NRG) technique with interleaving
($z$-averaging). We show that overbroadening errors can be largely
eliminated, that higher-moment spectral sum rules are satisfied to a good
accuracy, and that positions, heights and widths of spectral features are
well reproduced; the NRG approximates very well the spectral-weight
distribution. We find, however, that the discretization of the
conduction-band continuum nevertheless introduces artefacts. We present a
new discretization scheme which removes the band-edge discretization
artefacts of the conventional approach and significantly improves the
convergence to the continuum ($\Lambda\to1$) limit. Sample calculations of
spectral functions with high energy resolution are presented. We follow in
detail the emergence of the Kondo resonance in the Anderson impurity model
as the electron-electron repulsion is increased, and the emergence of the
phononic side peaks and the transition from the spin Kondo effect to the
charge Kondo effect in the Anderson-Holstein impurity model as the
electron-phonon coupling is increased. We also compute the spectral function
of the Hubbard model within the dynamical mean-field theory (DMFT),
confirming the presence of fine structure in the Hubbard bands.
\end{abstract}

\maketitle

\newcommand{\expv}[1]{\left\langle #1 \right\rangle}
\newcommand{\korr}[1]{\langle\langle #1 \rangle\rangle}

\renewcommand{\Im}{\mathrm{Im}}
\renewcommand{\Re}{\mathrm{Re}}

\section{Introduction}

Condensed-matter systems often exhibit rather complex behavior due to strong
Coulomb repulsion between the electrons at short distances. These effects
become very pronounced when electrons are strongly confined either in inner
electron shells (transition and rare-earth atoms) or in artificial
nanostructures (quantum dots). Theoretical studies of the corresponding
many-particle problems rely increasingly on advanced computational
techniques such as the numerical renormalization group (NRG)
\cite{wilson1975, krishna1980a, bulla2008}. The NRG allows to study both
static and dynamic \cite{sakai1989, sakai1992, costi1994, bulla1998,
hofstetter2000, peters2006, weichselbaum2007} properties of quantum impurity
models like the Kondo model or the Anderson impurity model. Applications
range from studies of thermodynamic properties of magnetic impurities in
normal \cite{krishna1980a, krishna1980b, oliveira1981} and superconducting
\cite{satori1992, sakai1993} host metals, dissipative two-state systems
\cite{costi1996akm}, electron transport through nanostructures
\cite{izumida1998}, to the use of the NRG as an impurity solver in the
dynamical mean-field theory (DMFT) \cite{sakai1994, bulla1999,
pruschke2000, georges1996}.

The foundation of the NRG is the transformation of a model with an infinite
number of degrees of freedom (the continuum of the conduction-band electron
states) to a model with a finite number of lattice sites (known as the
``hopping Hamiltonian'' or the ``Wilson chain'') which is numerically
tractable using a computer. This transformation consists of three steps: 1)
logarithmic discretization of the conduction band into increasingly narrow
intervals around the Fermi level, 2) dismissal of combinations of states
which do not couple directly to the impurity, and 3) unitary transformation
to a basis in which the conduction-band Hamiltonian takes the form of a
semi-infinite chain with exponentially decreasing hopping between
neighboring sites. In the first step, the discretization is controlled by a
parameter $\Lambda>1$, which sets the energy widths $\sim \Lambda^{-n}$ of
the intervals; the continuum is restored in the $\Lambda \to 1$ limit, while
typical values used in practical calculations are $\Lambda=2$ or even much
higher, depending on the application. The main approximation in the NRG
intervenes in the second step (dismissal of higher modes); this
approximation is controlled and it becomes better as $\Lambda$ is decreased
\cite{wilson1975}. An alternative discretization scheme \cite{campo2005}
leads directly to the decoupling of higher modes at the price of using a
non-orthogonal basis. The third step (mapping from the ``star Hamiltonian''
to a ``chain Hamiltonian'') can, in fact, be omitted \cite{bulla2005} at the
cost of significantly higher computational requirements. 

After these initial steps, the Hamiltonian is diagonalized iteratively,
taking one more chain site into account in each NRG iteration. Since the
Hilbert space grows exponentially, only a finite number of low-lying states
are kept in each iteration, while high-energy states are discarded
(truncated). This procedure is possible due to the ``separation of energy
scales'' which simply means that the matrix elements between the bottom and
top end of the excitation spectrum are small \cite{wilson1975}; this is an
important property of quantum impurity models. Truncation is another source
of systematic errors in NRG. These errors are more difficult to estimate
a-priori, but they can be kept small by a proper choice of $\Lambda$ and by
performing the truncation at suitably high cutoff energy.

While the NRG is the method of choice to study low-energy properties of
quantum impurity models, it is, however, commonly believed that it has
inherently limited energy resolution at higher energies due to the 
discretization of the conduction band. This is particularly relevant for the
calculations of dynamic properties \cite{sakai1989, sakai1992, costi1994},
such as the impurity spectral function or the dynamical susceptibilities.
Since the continuum impurity model is mapped onto a finite chain, the
spectral function consists of a set of delta peaks with given energies and
weights. These peaks need to be broadened \cite{costi1994, bulla2001,
bulla2008} to obtain the desired final result: a smooth spectral density
function. In order to efficiently smooth out spurious oscillations,
broadening kernel functions with long tails are usually chosen. The
log-Gaussian broadening function $\exp\left(-(\ln\omega -
\ln\omega')^2/b^2\right)$ is very commonly used since it is well adapted to
the logarithmic discretization grid. Unfortunately, the slowly decaying
tails lead to strong overbroadening effects, restraining the effective
energy resolution at higher energies and completely washing out any narrow
spectral features with small spectral weight.

Narrower broadening functions can be used when the so-called interleaved
method (also known as the ``$z$-averaging'') is used \cite{oliveira1994,
silva1996, paula1999, campo2005}. The interleaved method consists of
performing several NRG calculations for different (interleaved) logarithmic
discretization meshes controlled by the ``twist'' parameter $z \in (0:1]$.
In this way, the information is sampled from different energy regions in
each NRG run. The spectral function is then computed by averaging over all
$z$ values. Although the interleaved method does not truly restore the
continuum $\Lambda \to 1$ limit, it is surprisingly successful in removing
oscillatory features in the spectra; even averaging over only two values of
$z$ is often very beneficial. 

In this work, we study to what extent the energy resolution of the NRG can
be ultimately improved by the interleaved method. We perform the averaging
over a very large number of values of $z$ and use very narrow Gaussian
broadening kernel of width proportional to the energy of each individual
delta peak. This approach, although rather costly in terms of the required
computational resources, eliminates overbroadening and provides spectral
functions with very high energy resolution even on the energy scale of the
width of the conduction band. In addition to allowing us to study the fine
structure in the spectral functions of impurity models, this high-resolution
approach also uncovers the artefacts which are inherent in the NRG and
cannot be eliminated by the $z$-averaging. The artefacts diminish as
$\Lambda$ is decreased, but they are present in any practical NRG
calculation. By determining the appearance of the artefacts and their
expected locations, one can properly take them into account when
interpreting the results. We also propose a new discretization procedure
which is very successful in removing the most severe NRG discretization
artefacts. This improvement makes NRG a powerful technique for accurately
studying both low and high energy scales, thereby increasing its value as a
reliable impurity solver in DMFT.

This work is structured as follows. We introduce the Anderson impurity model
in Sec.~\ref{secmodel} and the details of the NRG calculations in
Sec.~\ref{secnrg}. To explore how accurately NRG approximates the
spectral-weight distribution, we present in Sec.~\ref{secsumrule} the sum
rules for spectral functions of the Anderson impurity model, the fulfilment
of which is then studied in Sec.~\ref{secnumeric}. The discretization
artefacts are discussed in Sec.~\ref{artefacts}, while in
Sec.~\ref{secremoval}, we present the modification to the discretization
scheme which renders these artefacts less severe. In Sec.~\ref{secexample}
we present examples of high-resolution spectral functions for the Anderson
and Anderson-Holstein impurity models which reveal interesting details,
which cannot be easily obtained by any other method. Finally, in
Sec.~\ref{secdmft} we demonstrate the feasibility of using the
high-resolution NRG approach in a DMFT setup. The resolution is sufficient
to resolve the fine structure in the Hubbard bands, in particular the
accumulation of the spectral weight at inner Hubbard band edges.

\section{Anderson impurity model}
\label{secmodel}

We consider the Anderson impurity model \cite{anderson1961}, the paradigm of
the quantum impurity models. It is defined by the following Hamiltonian
\begin{equation}
\begin{split}
H= &\sum_{k\sigma} \epsilon_{k} c^\dag_{k\sigma} c_{k\sigma}
+ \epsilon\, n + U n_\uparrow n_\downarrow \\
+ & \frac{1}{\sqrt{N}} \sum_{k\sigma} V_k 
\left( c^\dag_{k\sigma} d_{\sigma} + d^\dag_{\sigma} c_{k\sigma} \right),
\end{split}
\end{equation}
where operators $c_{k\sigma}$ describe the continuum conduction-band
electrons and operators $d_\sigma$ the impurity level, $\epsilon_k$ is the
band dispersion, $V_k$ the impurity hybridisation, $N$ the number of the
lattice sites, $\epsilon$ the impurity energy and $U$ the on-site
electron-electron repulsion. Furthermore, $n_\sigma=d^\dag_\sigma d_\sigma$
and $n=n_\uparrow + n_\downarrow$. In the derivations to follow, it is more
convenient to rewrite the hybridisation part of the Hamiltonian as
\cite{wilson1975}
\begin{equation}
H_\mathrm{hyb} = 
V \sum_\sigma 
\left( 
f^\dag_{0\sigma} d_\sigma + d^\dag_\sigma f_{0\sigma} 
\right).
\end{equation}
Here the hybridisation constant $V$ is defined as
\begin{equation}
V^2 = \frac{1}{N} \sum_k |V_k|^2
\end{equation}
and the operator $f_{0\sigma}$ as
\begin{equation}
f_{0\sigma} = \frac{1}{\sqrt{N}} \sum_k \frac{V_k}{V} c_{k\sigma}.
\end{equation}
The operator $f_{0\sigma}$ thus describes the combination of band states
which couple directly to the impurity level. The hybridisation strength is
given by $\Gamma=\pi \rho V^2$, where $\rho$ is the density of states (DOS)
in the conduction band. In numerical calculations we will use a constant DOS
$\rho=1/2D$, where $2D$ is the bandwidth, unless noted otherwise.

In the NRG, the continuum of band electrons is reduced to the hopping
Hamiltonian
\begin{equation}
H_{\mathrm{band}}^{\mathrm{(NRG)}} =
\sum_{i=0,\sigma}^{\infty} t_i 
\left( f^{\dag}_{i,\sigma} f_{i+1,\sigma}
+\text{H.c.} \right).
\end{equation}
The operator $f_{0\sigma}$ represents the previously introduced combination
of states, while $f_{i\sigma}$ for $i \geq 1$ describe further orbitals
along the Wilson chain. The coefficients $t_i$ depend on the discretization
scheme and on the parameters $\Lambda$ and $z$; asymptotically they behave
as $t_i \sim \Lambda^{-i/2}$. We emphasize that this is not an exact
representation of the continuum band Hamiltonian.

\section{Method}
\label{secnrg}

Dynamical NRG calculations were performed using the density-matrix approach
\cite{hofstetter2000, sidecoupled, toth2008nrg} using the density matrix
computed at the energy scale of $10^{-12}D$. Spectral functions were
obtained by delta-peak broadening using a Gaussian kernel with a width
proportional to the peak energy \cite{sakai1989, costi1991}:
\begin{equation}
\label{kernel}
P(\omega, E) = \frac{1}{\sqrt{2\pi}\eta_E}
e^{-\frac{(\omega-E)^2}{2\eta_E^2}},
\end{equation}
where $\omega$ is the energy of the point in the spectrum, $E$ is the
delta-peak energy and the width of the Gaussian is $\eta_E = \eta |E|$ with
$\eta$ a constant (we mostly use $\eta=0.01$ or $\eta=0.015$); the relative
spectral resolution is thus expected to be constant, $\Delta E/E \approx
\eta$. For the purposes of obtaining high-resolution spectral functions, it
is very important to use Gaussian broadening rather than, for example,
Lorentzian broadening, due to the fast decrease to zero of the Gaussian
function. We also note that the conventional log-Gaussian broadening kernel
$\exp\left(-(\ln\omega - \ln\omega')^2/b^2\right)$ becomes equivalent to a
simple Gaussian kernel for small enough $b$, aside from a small asymmetry of
the log-Gaussian function. Furthermore, parameters $\eta$ and $b$ are
related by $b=\sqrt{2}\eta$ in this limit. Nevertheless, the symmetry of the
Gaussian function is beneficial for the purposes of this work. For some
further comments on the spectral function broadening, see Appendix A.

The discretization was performed using the non-orthogonal-basis-set approach
of Campo and Oliveira \cite{campo2005}, with averaging over $N_z=32$ or
$N_z=64$ values of the twist parameter $z$, equally distributed in the
interval $(0:1]$. We note that in order to obtain a smooth spectrum, $\eta$
and $N_z$ need to be chosen such that $\eta N_z$ is of order 1.

The truncations were performed at an energy cutoff $E_\mathrm{cutoff} =
10\omega_N$, where $\omega_N \propto \Lambda^{-N/2}$ is the characteristic
energy scale at the $N$-th NRG iteration. When necessary, additional states
were retained above this cutoff energy to ensure that the truncation was
performed within an energy ``gap'' of at least $0.01 \omega_N$, so as not to
introduce systematic errors which may arise by retaining only parts of
clusters of nearly degenerate states. Charge conservation and
$\mathrm{SU}(2)$ spin invariance have been explicitly taken into account.

Spectral functions were obtained by ``patching'' together spectral functions
from every second energy shell (the $N/N+2$ approach) \cite{bulla2001}. The
details of the patching approach are important and, if not done properly,
the procedure will accentuate the discretization artefacts. At every
even-$N$ NRG interaction, we perform the patching as described in
Ref.~\onlinecite{bulla2001}: we merge spectral peaks in the energy range $[p
\omega_N:p \Lambda \omega_N]$ (unmodified) and spectral peaks in the range
$[p \Lambda \omega_N: p \Lambda^2 \omega_N]$ (after linear rescaling) with
the total spectral density; $p$ is some constant that we refer to as the
``patching parameter''. We return to the patching procedure in
Sec.~\ref{artefacts}, where we also comment on the relative merits of the
patching approach and the complete-Fock-space technique \cite{peters2006,
weichselbaum2007}.

\section{Higher-moment spectral sum rules for the Anderson impurity model}
\label{secsumrule}

A simple way of quantifying the distribution of the spectral weight is
through the moments, defined as
\begin{equation}
\mu_m = \int_{-\infty}^{\infty} \omega^m A_\sigma(\omega) \mathrm{d}\omega.
\end{equation}
where $A_\sigma(\omega)=-\frac{1}{\pi} \Im \korr{d_\sigma;
d^\dag_\sigma}_\omega$ is the spectral function. A stringent test of the
calculated {\sl dynamic} property (spectral function) is to verify that it
satisfies the sum rules which relate the moments to various {\sl static}
quantities (expectation values). The zero-th moment is simply the
normalization condition for spectral functions
\begin{equation}
\mu_0 = 1.
\end{equation}
Higher-moment spectral sum rules for the Anderson impurity model can be
derived as \cite{bulla2008, white1991moments}
\begin{equation}
\label{sumrule}
\mu_m = \expv{ \left\{ \left[ d_\sigma,H \right]_m, d^\dag_{\sigma}
\right\} },
\end{equation}
where $[A,B]_m$ is the iterated commutator, defined recursively as 
\begin{equation}
\begin{split}
[A, B]_1 &= [A,B]=AB-BA \\
[A, B]_{n+1} &= [[A,B]_n,B]
\end{split}
\end{equation}
while $\{A,B\}=AB+BA$ is the anticommutator. The first moment (mean energy)
is simply the Hartree energy of the impurity level,
\begin{equation}
\mu_1=\epsilon+U \expv{n_{-\sigma}},
\end{equation}
while the second is
\begin{equation}
\mu_2=V^2+\epsilon^2+(U+2\epsilon)U \expv{n_{-\sigma}}.
\end{equation}
The variance of the spectral function is thus
\begin{equation}
\kappa_2=\mu_2-\mu_1^2 = V^2 + U^2 \expv{n_{-\sigma}} (1-\expv{n_{-\sigma}}),
\end{equation}
i.e. a sum of the hybridisation width $V^2=\Gamma/(\pi\rho)$ and the
interaction-induced width. The third moment is
\begin{equation}
\begin{split}
\mu_3 =& \epsilon^3+2\epsilon V^2 + U(3\epsilon^2+3\epsilon U
+U^2+4V^2) \expv{n_{-\sigma}} \\
&- \frac{UV}{2} 
\left( 4 V \expv{n_{f,-\sigma}} + (U+2\epsilon) \expv{h^{(0)}_{-\sigma}} 
\right) \\
&+t_0 U V \expv{h^{(1)}_{-\sigma}}.
\end{split}
\end{equation}
Here the operator $n_{f,\sigma}$ is the $f_0$-orbital occupancy
$n_{f,\sigma}=f^\dag_{0\sigma} f_{0\sigma}$ and the operators $h$ are
hopping operators $h^{(i)}_\sigma= d^\dag_\sigma f_{i,\sigma} +
f^\dag_{i,\sigma} d_\sigma$ between the impurity orbital and the site $i$ of
the Wilson chain. The third central moment is thus
\begin{equation}
\begin{split}
\kappa_3 &= \mu_3-3\mu_1\mu_2+2\mu_1^3 = \\
&
U^3 \left( 2\expv{n_{-\sigma}}^3-3 \expv{n_{-\sigma}}^2+
\expv{n_{-\sigma}} \right) \\
&-V^2 \left( \epsilon+U \left( 2\expv{n_{f,-\sigma}}-\expv{n_{-\sigma}} 
\right) \right) \\
&-\frac{U V (U+2\epsilon)}{2} \expv{h^{(0)}_{-\sigma}} + 
t_0 U V \expv{h^{(1)}_{-\sigma}},
\end{split}
\end{equation}
which simplifies in the non-interacting limit to $\kappa_3 =-\epsilon V^2$.

The fourth moment is
\begin{widetext}
\begin{equation}
\begin{split}
\mu_4 =& \epsilon^4+3\epsilon^2 V^2 +V^4
+U \left( 4\epsilon^3+6\epsilon^2 U+4\epsilon U^2+U^3
+2(7\epsilon+4U)V^2 \right) \expv{n_{-\sigma}} \\
&+\quad UV\left[ (U+2\epsilon)^2 \expv{h^{(0)}_{-\sigma}} +
V \left( \left( 8\epsilon+3U \right) \expv{n_{f,-\sigma}} + U \expv{g_{-\sigma}} \right)
\right]
+ t_0^2 V^2+2t_0U(U+2\epsilon) \expv{ h^{(1)}_{-\sigma} },
\end{split}
\end{equation}
\end{widetext}
where operator $g_\sigma=T+2 (O_\perp +n_\sigma n_{f_\sigma})$; here
$T=d^\dag_\uparrow d^\dag_\downarrow f_{0,\uparrow} f_{0,\downarrow}
+\text{h.c.}$ is the two-particle hopping operator and $O_\perp =
d^\uparrow_\uparrow d_\downarrow f^\dag_{0,\downarrow} f_{0,\uparrow} +
\text{h.c.}$ is the transverse part of the spin-exchange operator. In the
non-interacting limit, the fourth moment simplifies to
\begin{equation}
\mu_4 = \epsilon^4+(3\epsilon^2+t_0^2)V^2 + V^4.
\end{equation}

It is important to point out that the expressions for $\mu_3$ and $\mu_4$
depend on the discretization through the coefficient $t_0$ and the operator
$h^{(1)}_{-\sigma}$ (for $\mu_3$ this is the case only for $U \neq 0$). They
are therefore not exact. While it is possible to derive exact expressions in
terms of $V_k$, $\epsilon_k$ and operators $d^\dag_\sigma c_{k,\sigma}
+\text{H.c.}$, they are of little practical use. This implies that in the
interacting case, calculations of $\mu_3$ and $\mu_4$ and the fulfilment of
the corresponding sum rules must be considered above all as a test of the
internal consistency of the method and of the extent of errors brought about
by the NRG truncation (``energetics''). Comparison with exact $\mu_3$ and
$\mu_4$ (were they known) would inevitably show some discrepancy (in the
following, we will demonstrate such behavior for $\mu_4$ in the
non-interacting case).

\section{Spectral weight distribution and sum rules}
\label{secnumeric}

\subsection{Non-interacting case}

We first consider the non-interacting ($U=0$) resonant-level model. The
spectral moments are tabulated in Table~\ref{table2}. The spectral function
of this model is given exactly as
\begin{equation}
\label{exact}
A(\omega) = - \frac{1}{\pi} \Im 
\left( \frac{1}{\omega-\epsilon+\Delta(\omega)} \right)
\end{equation}
with 
\begin{equation}
\Delta(\omega)=\Gamma\left[
i+\frac{1}{\pi} \ln \left(\frac{1-\omega/D}
{1+\omega/D}\right) 
\right]
\end{equation}
for $\omega \in [-D,D]$. This expression for $A(\omega)$ is used to compute
the reference values for spectral moments exactly (second column,
$\mu_i^{(e)}$). The right-hand sides of the sum rules, Eq.~\eqref{sumrule},
are computed in the standard way with ${\bar \beta}=0.75$ (third column,
$\mu_i^{(s)}$) \cite{wilson1975, krishna1980a, bulla2008}.  The fourth
column contains moments calculated by summing the suitably weighted
delta-peak contributions to the spectral function, $\mu_i^{(d)}$, and
finally the fifth column contains moments calculated directly by performing
a numerical integration with a spectral function after broadening,
$\mu_i^{(b)}$.

\begin{table*}[htb]
\centering
\begin{ruledtabular}
\begin{tabular}{@{}cllll@{}}
Moment & Exact, $\mu_i^{(e)}$ & Static, $\mu_i^{(s)}$ &
Dynamic (delta peaks), $\mu_i^{(d)}$ & 
Dynamic (broadened), $\mu_i^{(b)}$ \\
\colrule
$\mu_0$ & 1           &             & 0.999442     &  0.999981   \\
$\mu_1$ & -0.050000   & -0.050000   & -0.049983    & -0.049999   \\
$\mu_2$ & 0.0056831   & 0.0056831   &  0.0056866   & 0.0056871   \\
$\mu_3$ & -0.00044331 & -0.00044331 & -0.00044366  & -0.00044389 \\
$\mu_4$ & 0.00110129  & 0.0010225   &  0.0010220   & 0.0010225   \\
\end{tabular}
\end{ruledtabular}
\caption{Moments for the non-interacting impurity model with parameters
$\epsilon/D=-0.05$ and $\Gamma/D=0.005$. NRG parameters are $\Lambda=2,
\eta=0.015, N_z=32, p=2$.
}
\label{table2}
\end{table*}

The first three moments calculated as static quantities, $\mu_i^{(s)}$,
trivially agree with exact values since they are constants, while there is a
7 percent discrepancy for the fourth. This can be attributed to the
discretization errors as described previously in Sec.~\ref{secsumrule}. It
must be noted, however, that the fourth moment of a Lorentzian peak located
near the Fermi level strongly depends on the details around the band edges
and contains little information about the spectral distribution in the
frequency range of interest (i.e. around the peak itself). More importantly,
we find good agreement between $\mu_i^{(s)}$ and the moments computed from
dynamic quantities, $\mu_i^{(d)}$ and $\mu_i^{(b)}$, with errors in the few
permil range. This internal self-consistency of the method implies that the
accuracy of the energy levels in the range where the contributions to the
spectral function are sampled from is very good. The difference between
results from a calculation from delta-peak weights, $\mu_i^{(d)}$, or from
broadened spectral function, $\mu_i^{(b)}$, is remarkably small. This
already suggest that the broadening procedure itself does not lead to any
appreciable overbroadening.

\begin{figure}[htbp]
\includegraphics[width=8cm,clip]{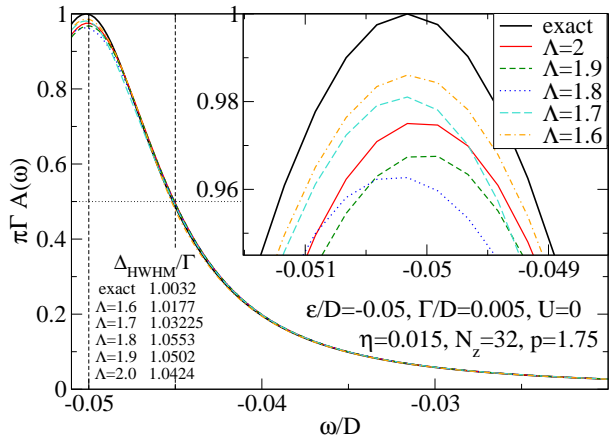}
\caption{(Color online) Spectral function of the non-interacting model for a
range of discretization parameters $\Lambda$, compared with
the exact solution, Eq.~\eqref{exact}.} \label{fig0}
\end{figure}

To study how the logarithmic discretization affects the spectral weight
distribution, we plot the spectral function of the non-interacting model for
a range of values of the discretization parameter $\Lambda$,
Fig.~\ref{fig0}. The peak position, width and height are well reproduced;
the position to within less than one percent even at $\Lambda=2$, while the
height and the half-width at half-maximum both deviate by less than 5
percent. As expected, the agreement improves as $\Lambda$ is decreased,
although not in a uniform manner. It may be noted that some spectral weight
seems to be missing in the peak (with the situation improving as $\Lambda
\to 1$). This is indeed the case; the missing spectral weight is located in
the NRG discretization artefacts that are the topic of Sec.~\ref{artefacts}.

\subsection{Interacting case}

We now switch on the interaction and consider an asymmetric Anderson
impurity model in the Kondo regime, $U/\pi\Gamma \gg 1$. Exact results for
moments are not available in this case, but we can compare $\mu_i^{(s)}$ and
$\mu_i^{(d)}$, Table~\ref{table4}. We find a similar degree of agreement
(few permil) as in the non-interacting case. We also observe that the
moments $\mu_i^{(d)}$ (and $\mu_i^{(b)}$) calculated for each value of $z$
separately depend relatively little on $z$. This is somewhat surprising
given that unaveraged spectral functions are extremely oscillatory. It also
implies that if we are really interested in a quantity which can be
expressed as an integral of the spectral function multiplied by some
relatively smooth weight function, there is only little benefit in
performing the $z$-averaging. 

\begin{table*}[htb]
\centering
\begin{ruledtabular}
\begin{tabular}{@{}clll@{}}
Moment & 
Static, $\mu_i^{(s)}$ & 
Dynamic (delta peaks), $\mu_i^{(d)}$ & 
Dynamic (broadened), $\mu_i^{(b)}$ \\
\colrule
$\mu_0$ &               &  1.000303     &   1.000306 \\
$\mu_1$ &  -0.0123204   &  -0.0123184   &   -0.0123184 \\
$\mu_2$ &  0.00455271   &  0.00455556   &   0.00455549 \\
$\mu_3$ &  -0.000138146 &  -0.000138222 &   -0.0001381970 \\
$\mu_4$ &  0.0010179    &  0.00101737   &   0.00101748 \\
\end{tabular}
\end{ruledtabular}
\caption{Moments for the asymmetric Anderson model
with parameters $U/D=0.07$, $\epsilon/D=-0.05$, $\Gamma/D=0.005$.
NRG parameters are $\Lambda=2$, $\eta=0.015$, $N_z=32$ and $p=2$.}
\label{table4}
\end{table*}

\begin{figure}[htbp]
\includegraphics[width=8cm,clip]{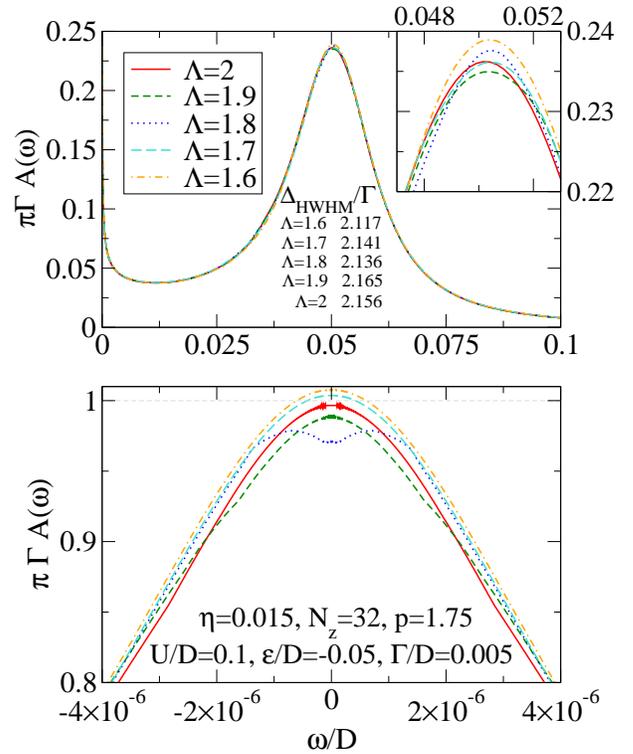}
\caption{(Color online) Spectral function of the symmetric Anderson impurity
model for a range of the discretization parameter $\Lambda$.  } \label{fig0a}
\end{figure}

We now study the spectral function of the symmetric Anderson impurity model
shown for a range of discretization parameters $\Lambda$ in
Fig.~\ref{fig0a}. The spectral functions overlap to a very good
approximation and there is little systematic overbroadening.  The width of
the charge-transfer peak is, as expected, approximately $2\Gamma$. The Kondo
resonance is well reproduced with a notable exception of $\Lambda=1.8$,
where we find an artefact which takes the form of a depression at the top of
the Kondo resonance. For this value of $\Lambda$, the Friedel sum rule
$A(\omega=0) = 1/\pi\Gamma$ is strongly violated. This is another
manifestation of the NRG artefacts that will be discussed in the following;
the result is improved by tuning the patching parameter $p$.

\begin{figure}[htbp]
\includegraphics[width=8cm,clip]{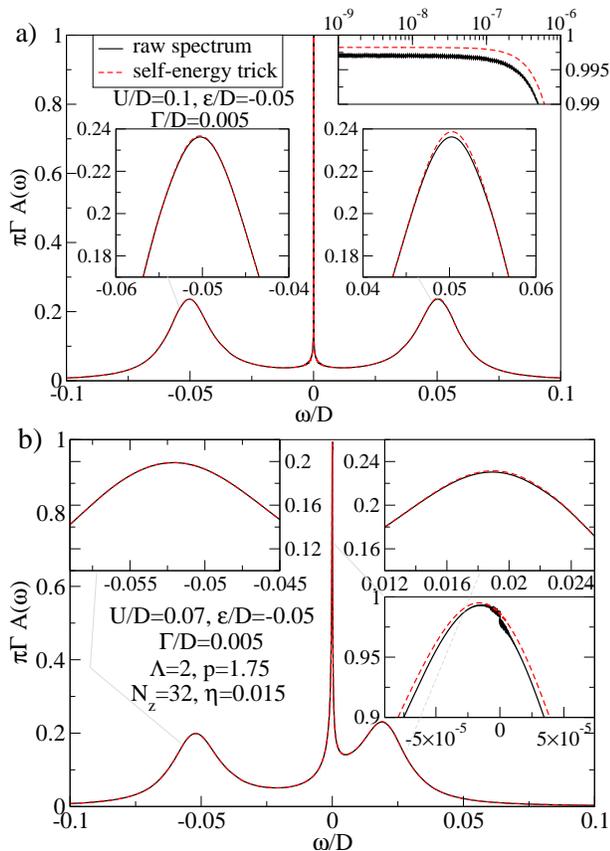}
\caption{(Color online) Spectral function of a) symmetric and b) asymmetric
Anderson impurity model: comparison of raw spectral function with that
obtained using the self-energy trick. } \label{fig5}
\end{figure}

A very successful method to reduce overbroadening effects in NRG
calculations is the ``self-energy trick'' \cite{bulla1998}. It consists of
numerically computing the self-energy as the ratio $\Sigma_\sigma(\omega)=U
F_\sigma(\omega)/G_\sigma(\omega)$ where $F_\sigma(\omega)= \langle\langle
n_{-\sigma} d_{\sigma};d^\dag_\sigma \rangle\rangle_\omega$ and
$G_\sigma(\omega) = \langle\langle d_{\sigma}; d^\dag_{\sigma}
\rangle\rangle_\omega$ and then computing an improved Green's function as
\begin{equation}
G^{\mathrm{improved}}_\sigma(\omega) = 
\frac{1}{\omega-\epsilon-\Sigma(\omega)+\Delta(\omega)}.
\end{equation}
An additional merit of this technique is that it leads to a partial
cancellation of the oscillatory features in $G_\sigma$ and $F_\sigma$,
giving a smooth self-energy $\Sigma_\sigma$. In Fig.~\ref{fig5} we compare
raw and self-energy-improved spectral functions for the symmetric and
asymmetric Anderson model. We first note that the change of the spectral
function upon using the self-energy trick is rather small, unlike in the
case of log-Gaussian broadening with large $b$ where the self-energy trick
leads to a sizable improvement and reduction of overbroadening. Results for
the symmetric case (Fig.~\ref{fig5}a) show that while the Friedel sum rule
is satisfied to better accuracy, the self-energy trick leads to slightly
broken particle-hole symmetry in the final result, which is not desirable.
On the other hand, in the general asymmetric case the self-energy trick
cures problems associated with different limiting behavior of $A(\omega)$
for $\omega \to 0^+$ and $\omega \to 0^-$, respectively (see
Fig.~\ref{fig5}b, inset with the close-up on the Kondo resonance).

\section{Discretization artefacts}
\label{artefacts}

\subsection{Types of artefacts}

Closer inspection of the computed high-resolution spectral functions reveals
the presence of artefacts which cannot be entirely eliminated by increasing
$N_z$ or reducing $\Lambda$. These are thus genuine intrinsic NRG
discretization artefacts.

\begin{figure*}[htbp]
\includegraphics[width=17cm,clip]{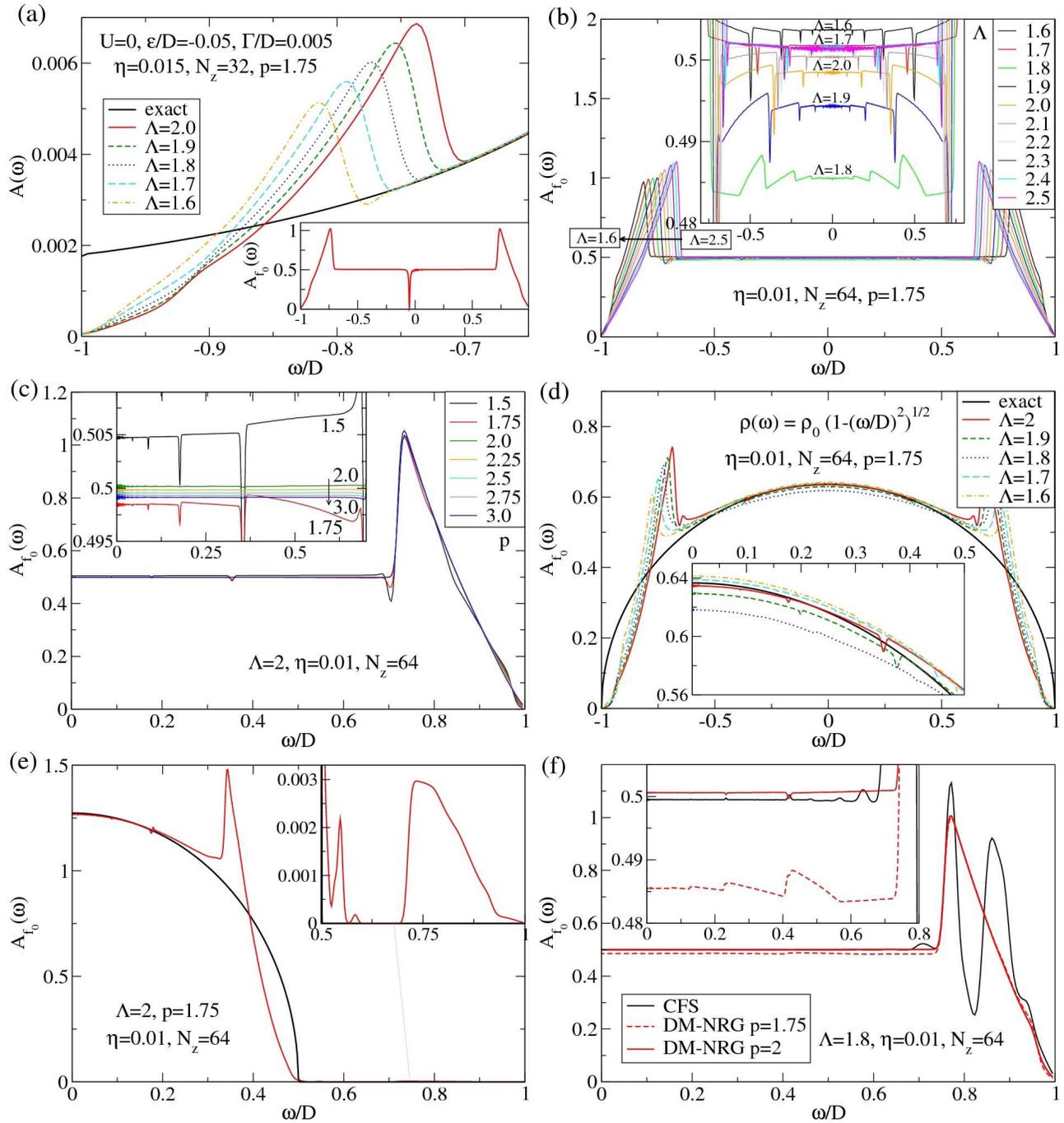}
\caption{(Color online) a) High-energy artefacts in the spectral function of
the resonant-level model. Inset: the spectral function on the first site of
the Wilson chain, $f_0$.
b) Spectral function on the first site of the free Wilson chain, $A_{f_0}$,
for different values of the discretization parameter $\Lambda$. 
c) Spectral function $A_{f_0}$ computed for different values of the spectral
patching parameter $p$.
d) Spectral function $A_{f_0}$ in the case of semi-elliptic DOS,
$\rho(\epsilon)=\rho_0 \sqrt{1-(\epsilon/D)^2}$.
e) Spectral function $A_{f_0}$ in the case of semi-elliptic DOS,
$\rho(\epsilon)=\rho_0 \sqrt{1-(2\epsilon/D)^2}$ with support $[-0.5:0.5]D$.
f) Comparison of spectral function $A_{f_0}$ computed with the
complete-Fock-space NRG approach and the conventional density-matrix NRG
approach.
} \label{fig6}
\end{figure*}

As a first example, we plot in Fig.~\ref{fig6}a the spectral function
$A(\omega)$ of the non-interacting impurity model in the high-energy range
near the band edge, i.e. the tail of the Lorentzian spectral peak. We see a
pronounced artefact which shifts toward the band-edge as $\Lambda$ is
decreased. If the exact solution is subtracted from the artefact, we find
that there is some cancellation of positive and negative differences, but
there is nevertheless a positive net (integrated) difference; this is the
origin of the previously mentioned missing spectral weight in the spectral
peak of the resonant-level model. In the inset we show the spectral function
$A_{f_0}(\omega)$ of the first site of the Wilson chain $f_0$. For a flat
band, $\rho=1/2D$, this function should likewise be flat, except for the
features that are mirrored from the impurity spectral function $A(\omega)$.
The NRG discretization, however, introduces additional artefact structure
for energies near the band-edge.

To study this problem more closely, we compute $A_{f_0}(\omega)$ for a
system without the impurity, Fig.~\ref{fig6}b. In addition to the very
pronounced band-edge artefact, there are also discernible additional
artefacts at lower energies. The ratio of energies for two consecutive
artefacts is $\Lambda$, as expected. The artefact peaks presumably exist
down to lowest energy scales, but their amplitudes decrease rapidly and
eventually the peaks can no longer be resolved since they are masked by the
residual oscillations in the calculated spectral functions. Curiously, the
average value of $A_{f_0}(\omega)$ in the low-energy region seems to have a
minimum for $\Lambda \approx 1.8$. Furthermore, for this value of $\Lambda$,
the artefacts appear to be the largest. This is in agreement with the
results for spectral functions presented above. These artefacts can,
however, be strongly reduced finding proper parameter $p$ of the spectrum
patching procedure.

In Fig.~\ref{fig6}c we plot the spectral density $A_{f_0}(\omega)$ for
different values of $p$. If $p$ is too small, we obtain very pronounced
discretization artefacts. If $p$ is too large, the spectral density is
underestimated. The optimal value of $p$ is around $2$, but it depends on
the energy cutoff in the truncation; we work with cutoff
$E_\mathrm{cutoff}=10\omega_N$, thus for $p=2$ and $\Lambda=2$, we have $p
\Lambda^2 \omega_N = 8 \omega_N < E_\mathrm{cutoff}$. We remark that the
large artefacts near the band-edge are not related to the patching procedure
(see also below), although the form of the artefacts does depend somewhat on
the value of $p$.

We can formulate the following recipe for choosing appropriate NRG
parameters:
\begin{enumerate}
\item fix $\Lambda$;
\item increase truncation cutoff until NRG results no longer change
significantly; 
\item tune $\eta$ and $N_z$ to suppress overbroadening of spectral
functions;
\item tune $p$ for good reproduction of the band spectral function
$A_{f_0}(\omega)$.
\end{enumerate}
If necessary, steps 2-4 may be reiterated. To be specific, for $\Lambda=2$,
$E_\mathrm{cutoff}=10\omega_N$, $\eta=0.01$, and $N_z=64$, we find that
$A_{f_0}(\omega)$ is closest to $1/2D$ at low energies for $p=2.1$. A caveat
is in order: tuning $p$ for good reproduction of $A_{f_0}(\omega)$ does not
necessarily imply that the same value of $p$ will be optimal for the full
problem (with the impurity coupled to the bath). Nevertheless, such $p$ is
most likely a good choice.

For applications of the NRG as an impurity solver in DMFT, it is important
to reproduce an arbitrary conduction-band DOS as accurately as possible. As
a simple test, in Fig.~\ref{fig6}d we consider the case of the cosine band
dispersion, $\epsilon_k=D\cos{k}$, which has a semi-elliptic DOS, 
\begin{equation}
\rho(\epsilon) = \rho_0 \sqrt{1-(\epsilon/D)^2}.
\end{equation}
We again find sizable artefacts near band edges at approximately the same
positions as in the case of a flat band. One might expect that using a
DOS with a limited support (such that it excludes the strong
artefacts at $\approx 0.7D$) would resolve the issue. Alas, that is not the
case. The artefacts simply appear at rescaled positions, as is shown in the
example of a semi-elliptic DOS with support $[-0.5:0.5]$,
Fig.~\ref{fig6}e. Any abrupt change in the density of states (any sharp
feature, in fact) is thus expected to lead to anomalies at low energies.

Spectra calculated using the complete-Fock-space (CFS) approach
\cite{peters2006, weichselbaum2007} also show artefacts, although there are
differences in the details, see Fig.~\ref{fig6}f. There are several
advantages to the CFS approach: the normalization is satisfied exactly
within numerical accuracy and there is no ambiguity in the choice of the
energy range where the spectrum is computed at each iteration (no parameter
$p$). The conventional approach is, however, significantly faster since the
eigenvectors and matrix elements need to be computed only in the retained
part of the Hilbert space in each NRG iteration. In addition, in CFS the
delta-peak energies are given as a difference between an energy of a kept
state and an energy of a discarded state; the latter is located at the upper
end of the shell excitation spectrum, thus it is affected by the accumulated
truncation errors from all previous NRG iterations. For this reason, the
spectra calculated using the traditional approach with patching satisfy
higher-moment sum rules to higher accuracy (in the permil range as opposed
to the percent range) even though they break the normalization sum rule.

\subsection{Origin of the band-edge artefacts}

\begin{figure}[htbp]
\centering
\includegraphics[width=8cm,clip]{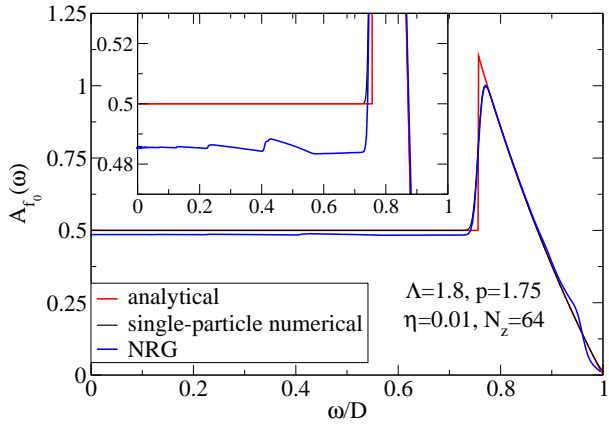}
\caption{Comparison of spectral functions $A_{f_0}$.}
\label{figan}
\end{figure}

In the case of a flat band, $\rho(\omega)=\mathrm{const}.$, the origin of
the main artefact near the band edge is easy to understand. Following
Ref.~\onlinecite{campo2005}, we write the density of states on site $f_0$ as
\newcommand{\Epsilon}{\mathcal{E}}
\begin{equation}
\label{cons}
A_{f_0}(\omega) = \frac{\epsilon_j^z-\epsilon_{j+1}^z}
{2D |d\Epsilon_j^z/dz|},
\end{equation}
where $\epsilon_j^z$ define the discretization mesh,
\begin{equation}
\begin{split}
\epsilon_1^z &= D, \\
\epsilon_j^z &= D \Lambda^{2-j-z} \quad (j=2,3,\ldots),
\end{split}
\end{equation}
and $\Epsilon_j^z$ are defined as
\begin{equation}
\label{Eps}
\Epsilon_j^z = \frac{\int_{I_j} d\epsilon}{\int_{I_j} d\epsilon/\epsilon}
= \frac{\epsilon_j^z-\epsilon_{j+1}^z}
{\ln\left( \epsilon_j^z/\epsilon_{j+1}^z \right)},
\end{equation}
with $I_j=[\epsilon_j^z;\epsilon_{j+1}^z]$, which gives
\begin{equation}
\begin{split}
\Epsilon_1^z &= D \frac{1-\Lambda^{-z}}{z \ln\Lambda}, \\
\Epsilon_j^z &= D \frac{1-\Lambda^{-1}}{\ln\Lambda} \Lambda^{2-j-z},
\quad (j=2,3,\ldots).
\end{split}
\end{equation}
For given argument $\omega$, the parameters $z$ and $j$ in the right hand
side of Eq.~\eqref{cons} are determined by the relation
$\Epsilon_j^z=\omega$ which has a unique solution. (To simplify the notation
and discussion, we assumed particle-hole symmetry of the conduction band and
we consider $\omega>0$ only. All features at positive energies are then
simply mirrored to negative frequencies.) It can be easily shown that for
$j=2,3,\ldots$, i.e. for 
\begin{equation}
\omega \in \left[ -\frac{1-\Lambda^{-1}}{\ln\Lambda};
+\frac{1-\Lambda^{-1}}{\ln\Lambda} \right],
\end{equation}
we indeed have 
\begin{equation}
A_{f_0}(\omega)=1/2D.
\end{equation}
This is not the case, however, for $j=1$, i.e. for $\omega$ within
$(1-\Lambda^{-1})/\ln\Lambda$ from the band edges. We obtain, instead,
\begin{equation}
A_{f_0}(\omega) = \frac{(1+\beta \omega)^2}
{\omega \left( \omega + \frac{1+\beta \omega}
{1-\omega^{\beta+1/\omega}} \right) \ln \Lambda} \\
\end{equation}
with $\beta = W \left[ -e^{-1/\omega}/\omega \right]$, where $W$ is the
Lambert W-function. In Fig.~\ref{figan} we plot three spectral functions: 1)
analytically calculated spectral function, $A^{(a)}_{f_0}$, 2) the spectral
function numerically calculated by exact diagonalisations of the
single-electron Hamiltonians obtained after discretization, $A^{(n)}_{f_0}$,
and 3) the spectral function calculated directly using NRG,
$A^{\mathrm{(NRG)}}_{f_0}$. Compared to the analytical result,
$A^{(a)}_{f_0}$, the function $A^{(n)}_{f_0}$ features artefacts due to
finite $N_z$ and broadening, while $A^{\mathrm{(NRG)}}_{f_0}$ in addition
shows truncation errors. The band-edge artefact is thus not some unexpected
numerical artefact, but it is the direct result of a particular choice of
the discretization scheme. It arises from a different behavior of
$\Epsilon^z_1$ as a function of $z$ as compared to other $\Epsilon^z_j$.
This, in turn, is due to the presence of the band-edge, which sets the upper
boundary in the integrals in Eq.~\eqref{Eps}.

\begin{figure}[htbp]
\centering
\includegraphics[width=8cm,clip]{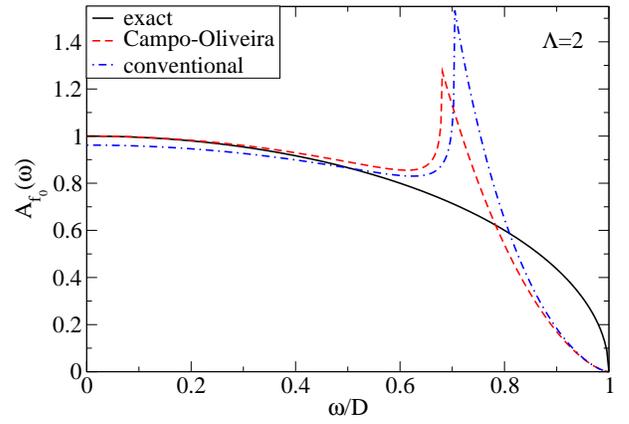}
\caption{Analytically computed spectral function $A_{f_0}$ for semi-elliptical
DOS in the conventional and Campo-Oliveira discretization scheme compared
with the exact DOS.}
\label{figx3}
\end{figure}

For arbitrary density of states we introduce weight functions for different
discretization intervals \cite{campo2005}
\begin{equation}
\phi_{j0} = \left( \frac{\rho(\epsilon)}{\int_{I_j} \rho(\omega) d\omega}
\right)^{1/2},
\end{equation}
so that the operator $f_0$ takes the following form
\begin{equation}
f_0 = \sum_j \left( \int_{I_j} \rho(\omega) d\omega \right)^{1/2} a_{j0},
\end{equation}
where $a_{jm}$ are conduction-band operators for the $m$-th mode
(combination of states) in the $j$-th discretization interval; only $m=0$
modes are retained in the NRG.
The spectral function on the first site of the Wilson chain is then
given as
\begin{equation}
\label{arb}
A_{f_0}(\omega) = \frac{\int_{I_j} \rho(\omega) d\omega}
{|d\Epsilon_j^z/dz|},
\end{equation}
where $z$ and $j$ are again determined by the relation
$\Epsilon_j^z=\omega$. In order to achieve decoupling of higher modes in
each discretization interval, Campo and Oliveira proposed to calculate
coefficients $\Epsilon_j^z$ as \cite{campo2005}
\begin{equation}
\Epsilon_j^z = \frac{\int_{I_j} \rho(\epsilon) d\epsilon}
{\int_{I_j} \rho(\epsilon)/\epsilon\, d\epsilon}.
\end{equation}
In the most commonly used conventional discretization scheme
\cite{yoshida1990}, the coefficients are given, instead, as
\begin{equation}
\Epsilon_j^z = \frac{\int_{I_j} \rho(\epsilon) \epsilon\, d\epsilon}
{\int_{I_j} \rho(\epsilon) d\epsilon}.
\end{equation}
It is easy to verify that $\Epsilon_j^z$ calculated in either way do not
satisfy the equation $A_{f_0}(\omega) = \rho(\omega)$ and that strong
artefacts appear near sharp features in the density of states. As an
example, we compare in Fig.~\ref{figx3} the cosine band DOS with $A_{f_0}$
computed with both discretization schemes. Both show significant band-edge
artefacts (see also Fig.~\ref{fig6}d). In the conventional scheme, the
spectral function $A_{f_0}$ in addition systematically underestimates
$\rho(\omega)$ at lower energy scales by the well-known factor of 
\begin{equation}
A_\Lambda = \frac{\ln\Lambda}{2} \frac{1+\Lambda^{-1}}{1-\Lambda^{-1}},
\end{equation}
which is taken into account in practical NRG calculations in an ad-hoc
manner by multiplying the impurity hybridisation (or exchange constant) by
this same value.

\section{Overcoming the discretization artefacts}
\label{secremoval}

We have demonstrated that the origin of the discretization artefacts is in
the $z$-dependence of the coefficients $\Epsilon_j^z$. These coefficients,
in turn, are determined by the discretization points $\epsilon_j^z$, the
choice of the basis states of the discretized conduction band (in particular
the weight functions $\phi_{j0}$) and the recipe for the calculation of
coefficients \cite{yoshida1990, campo2005}. Keeping the same set of the
discretization points and zero-mode basis states, we may decide to {\sl
define} $\Epsilon_j^z$ in a more appropriate way, i.e. such that all
coefficients satisfy the condition
\begin{equation}
\label{eq1}
\frac{\int_{I_j} \rho(\epsilon) d\epsilon}
{|d\Epsilon_j^z/dz|} = \rho(\omega).
\end{equation}
Details about solving this equation are given in the Appendix B.
Well-behaved solution may be found for arbitrary DOS function $\rho(\omega)$
and the asymptotic (large $j$) behavior of $\Epsilon_j^z$ is the same as in
the Campo-Oliveira discretization scheme.

We note that this modification of the discretization procedure in no way
makes NRG an exact method, even though we expect much better reproduction of
the conduction band DOS. In the spirit of the original NRG procedure, we
still rely on the assumption that discarding higher-mode states in each
discretization interval is a good approximation which can be systematically
improved by reducing $\Lambda$ toward 1. In particular,
discretization-related artefacts are still possible and we indeed find them,
as detailed in the following. The improvement consists in significantly
reducing the severity of the artefacts.

\begin{figure}[htbp]
\centering
\includegraphics[width=8cm,clip]{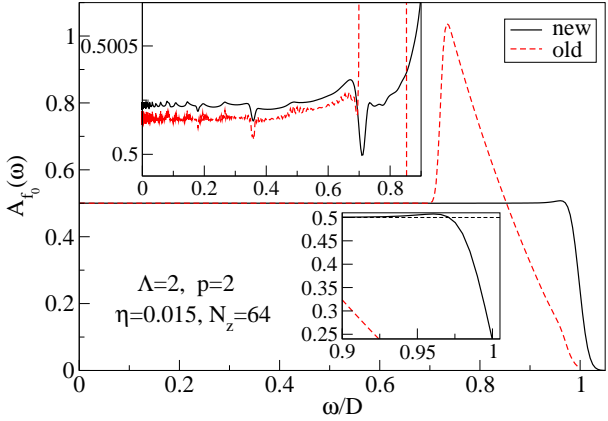}
\caption{Comparison between the Campo-Oliveira (old) and the improved
discretization (new) approaches.} \label{suc}
\end{figure}

Solving Eq.~\eqref{eq1} in the case of a flat band, only $\Epsilon_1^z$ is
modified, while $\Epsilon_j^z$ for $j \geq 2$ remain the same. We obtain
\begin{equation}
\Epsilon_1^z = \frac{1-\Lambda^{-z}}{\ln\Lambda} + 1-z.
\end{equation}
As $z$ is swept from 0 to 1, this quantity takes values over the same
interval as the Campo-Oliveira expression for $\Epsilon_1^z$. This is
important, since $\Epsilon_j^z$ must cover the whole energy range. In
Fig.~\ref{suc} we compare the spectral function $A_{f_0}(\omega)$ computed
with original and modified discretization approach. The improvement is, as
expected, significant. The spectral function overshoots slightly (by less
than two percent) as the band-edge is approached and it decays to zero on
the scale set by the broadening parameter $\eta$. A closer look reveals
small residual artefacts positioned at energies $\Epsilon_j^{z=1}$,
$j=1,2,\ldots$, which take the form of asymmetric dips. Their weights
rapidly decreases with increasing $j$; in the worst case, for $j=1$, the dip
amplitude is less than one permil of the background $1/2D$ weight. There are
further artefacts between the $\Epsilon_j^{z=1}$ dips, but their amplitudes
are even smaller than those of the main artefacts. At low energies,
$A_{f_0}(\omega)$ converges to $0.50025$, which can be tuned exactly to
$1/2$ by further tuning of the patching parameter $p$.

\begin{figure}[htbp]
\centering
\includegraphics[width=8cm,clip]{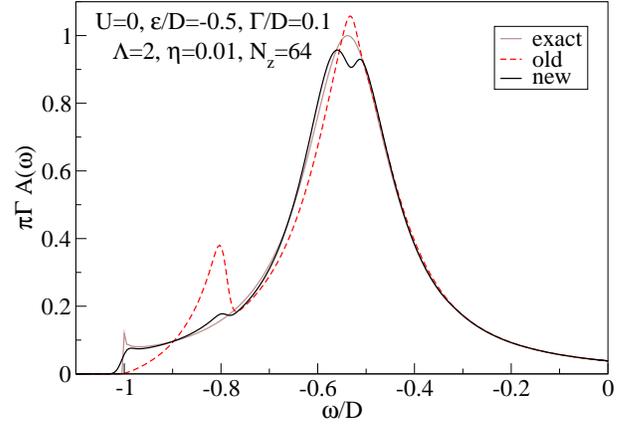}
\caption{Spectral function of the resonant-level model: comparison between
the exact analytical result and two different discretization approaches.}
\label{suc2}
\end{figure}

In Fig.~\ref{suc2} we compare the spectral functions of the resonant-level
model obtained using both discretization schemes. We find that the new
discretization scheme strongly suppresses the artefact peak structure and
correctly reproduces the behavior at the very edge of the conduction band
(within the limits imposed by the broadening procedure). We also see that
the flanks of the spectral peak agree better with the exact solution. On the
other hand, we see that an artefact appears at the very top of the
resonance. This artefact is directly connected with the discretization
itself and does not depend, for example, on the truncation or patching; the
situation improves, however, with decreasing $\Lambda$ (see
Fig.~\ref{convspec} in Subsection B below). We point out that the artefact
is not located at any $\Epsilon_j^{z=1}$, thus it is not related to the
residual artefacts found in $A_{f_0}(\omega)$ of the decoupled band. It
should rather be interpreted as a finite-size effect due to representation
of the continuum by a finite chain; the $z$-averaging cannot entirely
eliminate such effects. In spite of the artefact, we may conclude that the
overall reproduction of the spectral weight distribution is considerably
improved. It may also be noted that we present here the most difficult case:
a very broad resonance near the band edge. Such situation is rather unusual
for impurity problems; for narrow resonances and for peak energies closer to
the Fermi level the double-peak artefact is quickly reduced. Broad spectral
distributions are, however, typical for DMFT applications, where residual
artefacts may become more problematic.

In Tables~\ref{table2new} and \ref{table4new} we show the moments for the
non-interacting model and for the asymmetric Anderson model. They are to be
compared with the corresponding Tables~\ref{table2} and \ref{table4}. The
agreement between $\mu_i^{(s)}$ and $\mu_i^{(d)}$ in the new scheme is below
one permil for all moments, as in the old one. In the resonant-level model,
the agreement of calculated $\mu_4$ now agrees with the exact value within
one permil (while in the Campo-Oliveira scheme we found a discrepancy of
$7\%$). In the Anderson model, we also observe a change in $\mu_4$ of the
same order, suggesting a similar degree of improvement.

\begin{table*}[htb]
\centering
\begin{ruledtabular}
\begin{tabular}{@{}cllll@{}}
Moment & Exact, $\mu_i^{(e)}$ & Static, $\mu_i^{(s)}$ &
Dynamic (delta peaks), $\mu_i^{(d)}$  &
Dynamic (broadened), $\mu_i^{(b)}$ \\
\colrule
$\mu_0$ & 1           &             & 0.999979     & 0.999964    \\
$\mu_1$ & -0.050000   & -0.050000   & -0.049998    & -0.049997   \\
$\mu_2$ & 0.0056831   & 0.0056831   &  0.0056876   & 0.0056704   \\
$\mu_3$ & -0.00044331 & -0.00044331 & -0.00044376  & -0.00044217 \\
$\mu_4$ & 0.00110129  & 0.00110120  &  0.00110158   & 0.0010842   \\
\end{tabular}
\end{ruledtabular}
\caption{Moments for the non-interacting impurity model with parameters
$\epsilon/D=-0.05$ and $\Gamma/D=0.005$. Improved discretization
scheme, $\Lambda=2$, $\eta=0.015$, $N_z=32$, $p=2$.
}
\label{table2new}
\end{table*}

\begin{table*}[htb]
\centering
\begin{ruledtabular}
\begin{tabular}{@{}clll@{}}
Moment & 
Static, $\mu_i^{(s)}$ & 
Dynamic (delta peaks), $\mu_i^{(d)}$ & 
Dynamic (broadened), $\mu_i^{(b)}$ \\
\colrule
$\mu_1$ &               &  1.000302     &   1.000287 \\
$\mu_1$ &  -0.0123213   &  -0.0123193   &   -0.0123187 \\
$\mu_2$ &  0.00455274   &  0.00455661   &   0.0045386 \\
$\mu_3$ &  -0.000138138 &  -0.000138191 &   -0.000137434 \\
$\mu_4$ &  0.00109664    &  0.00109694   &   0.00107882 \\
\end{tabular}
\end{ruledtabular}
\caption{Moments for the asymmetric Anderson model
with parameters $U/D=0.07$, $\epsilon/D=-0.05$, $\Gamma/D=0.005$.
Improved discretization scheme, $\Lambda=2$, $\eta=0.015$, $N_z=32$, $p=2$.}
\label{table4new}
\end{table*}

\subsection{Arbitrary density of states}

\begin{figure}[htbp]
\centering
\includegraphics[width=8cm,clip]{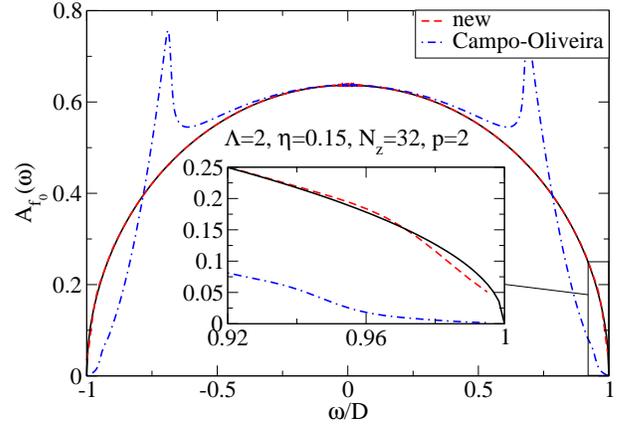}
\caption{Spectral function $A_{f_0}$ in the case of semi-elliptic
DOS computed using the new discretization scheme.}
\label{works}
\end{figure}

In Fig.~\ref{works} we demonstrate on the example of the semi-elliptic DOS
that the proposed discretization approach can also be applied for an
arbitrary density of states. In this case, all $\Epsilon_j^z$ are modified
and they need to be numerically calculated using the technique described in
the Appendix B. As in the case of flat band, some small discrepancies
between $A_{f_0}(\omega)$ and $\rho(\omega)$ are found at the very edge of
the band. The over-all agreement is, however, significantly improved on all
energy scales.

\begin{figure}[htbp]
\centering
\includegraphics[width=8cm,clip]{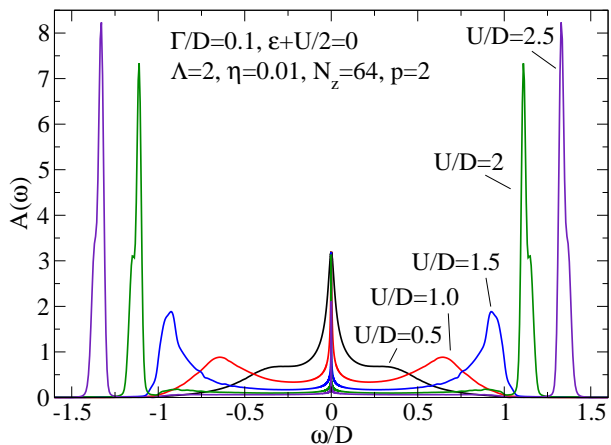}
\caption{Spectral function of the symmetric Anderson model with
semi-elliptic DOS computed using the new discretization scheme.}
\label{gamma01}
\end{figure}

We test the method on the case of a symmetric Anderson model with
semi-elliptic DOS. In Fig.~\ref{gamma01} we plot spectral functions for
rather large $\Gamma=0.1 D$ for increasing values of $U$. While for small
$U$ the functions are rather smooth, we observe more pronounced residual
artefacts for large values of $U$, as the charge-transfer (Hubbard) peaks
approach the band edge (see, for example, the $U/D=1.5$ case). Nevertheless,
the results are significantly more physically sensible than those obtained
using conventional broadening and discretization techniques. For $U \gtrsim
2D$, the Hubbard peaks are located outside the conduction band. They become
narrower and they have strongly asymmetric shape \cite{raas2005}; in fact,
in some parameter ranges they have a two-peak structure. We also note that
the impurity parameters used here are comparable to those that typically
arise in effective models in DMFT (see also Sec.~\ref{secdmft}).

\subsection{Convergence with $\Lambda$}

\begin{figure}[htbp]
\centering
\includegraphics[width=8cm,clip]{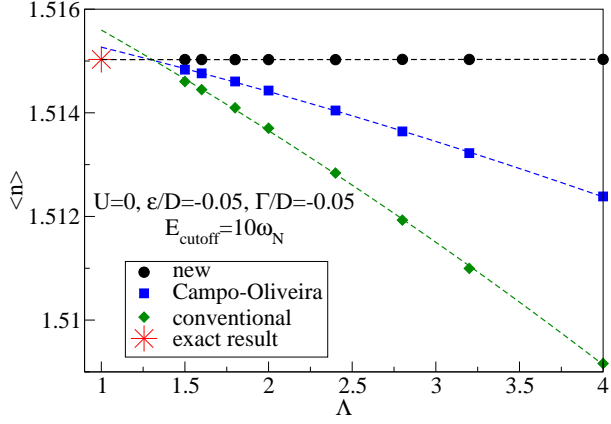}
\caption{Convergence of the expectation value of the level occupancy in the
resonant-level model with decreasing $\Lambda$ in different discretization
schemes. Dashed lines are fits with quadratic functions which serve to
perform a $\Lambda \to 1$ extrapolation.}
\label{figx1}
\end{figure}

The new discretization scheme vastly improves the convergence to the
$\Lambda \to 1$ limit. We demonstrate this in Fig.~\ref{figx1} by comparing
the calculated level occupancy in the resonant-level model with the exact
value as a function of $\Lambda$. With the new approach, we obtain very
accurate results even with very large discretization parameter (four digits
of accuracy at $\Lambda=8$). In other approaches, not only is the
convergence to the continuum limit slower, but extrapolating the numeric
results in the range $\Lambda \geq 1.5$ to the $\Lambda\to 1$ limit leads to
a systematic error; presumably the assumption of quadratic (or polynomial)
$\Lambda$-dependence no longer holds for smaller $\Lambda$. With the
improved discretization approach one can compute expectation values of
various operators reliably even at very large $\Lambda$: this is quite
important for numerically demanding multi-orbital/multi-channel quantum
impurity problems. Similar improvements also hold for calculations of
thermodynamic quantities (such as the impurity contribution to the magnetic
susceptibility and entropy).

\begin{figure}[htbp]
\centering
\includegraphics[width=8cm,clip]{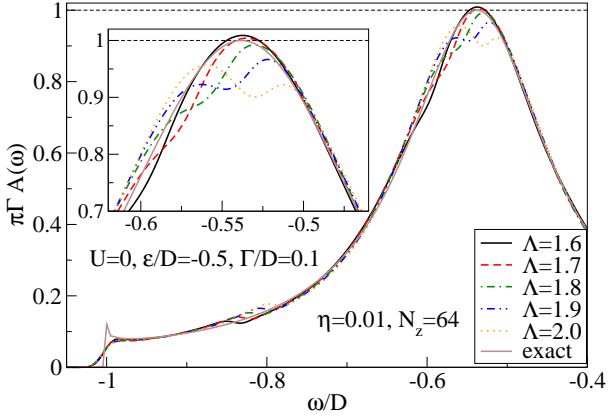}
\caption{Convergence of the spectral function of the resonant-level
model to the exact result with decreasing $\Lambda$ in the case
of the new discretization scheme.}
\label{convspec}
\end{figure}

We have seen previously that residual artefacts are observed in spectral
functions. In Fig.~\ref{convspec} we report how these residual artefacts are
reduced as $\Lambda$ is reduced. For sufficiently small $\Lambda$, the
artefact appearing as double peak structure is eliminated. Furthermore, we
see that the artefacts shift as a function of $\Lambda$. This implies that
some additional improvement could be obtained by performing the calculation
for several different values of $\Lambda$ and averaging the resulting
spectral functions.

In the sense that the new discretization scheme gives the best possible
representation of the conduction band DOS by the Wilson chain (after the
$z$-averaging), this technique provides the best results that one can
achieve by representing each discretization interval by a single level. A
possible systematic improvement would consist in including more than one
mode for low $j$ (where the band DOS still varies strongly as a function of
energy) and performing the NRG in the star basis, perhaps using Lanczos
exact diagonalisation procedure to diagonalize the cluster at each NRG
iteration.

\subsection{Spectral features outside the conduction band}

\begin{figure}[htbp]
\centering
\includegraphics[width=8cm,clip]{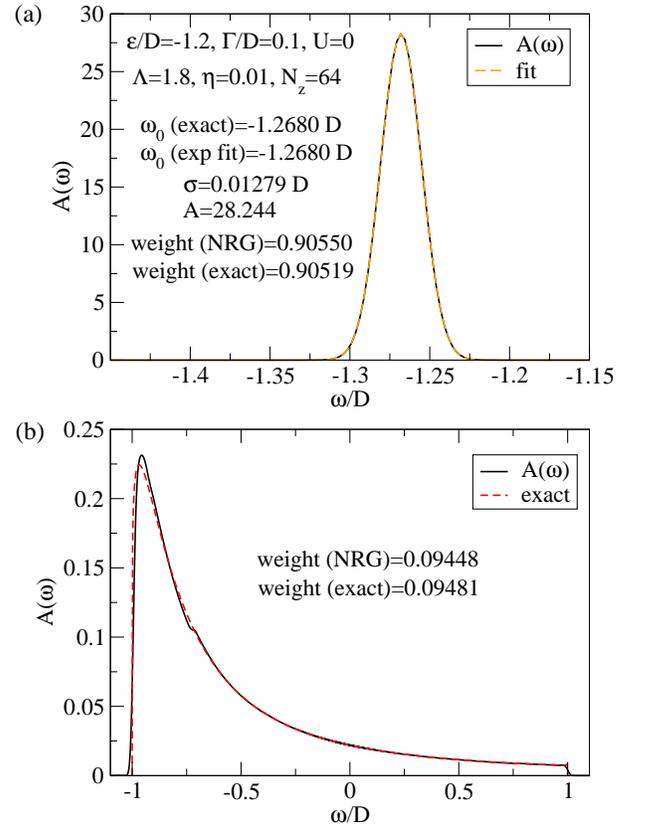}
\caption{Spectral function of the resonant-level model in the case where
the resonance is outside the conduction band.}
\label{fig55}
\end{figure}

We tested how accurately the NRG reproduces spectral features at energies
outside the energy band (i.e. outside the $[-D:D]$ interval in the case of a
flat conduction band) for the example of the resonant-level model. In this
model, for $\epsilon \lesssim -D$, there is a $\delta$-peak at the energy
$\omega_0$ given by
\begin{equation}
\omega_0 - \epsilon + \mathrm{Re} \Delta(\omega_0) = 0,
\end{equation}
with weight
\begin{equation}
\frac{1}{1+
\left( \frac{\partial \mathrm{Re} \Delta(\omega)}{\partial \omega}
\right)_{\omega=\omega_0}},
\end{equation}
while the spectrum in the $[-D:D]$ range is described by Eq.~\eqref{exact}.
We compare the calculated spectrum with the expected results in Fig.~
\ref{fig55}. The $\delta$-peak takes the form of the broadening kernel,
Eq.~\eqref{kernel}, and we can accurately extract its position, height and
width by fitting to an exponential function $A \exp[
-(\omega-\omega_0)^2/2\sigma^2 ]$. We find that the position and the (integrated)
weight of the peak are reproduced within approximately four digits of
precision. Furthermore, we find that 
\begin{equation}
\sigma/\omega_0 = 0.01009,
\end{equation}
which is to be compared with the broadening factor $\eta=0.01$. We conclude
that within one percent accuracy, there is no other source of broadening
than the explicit spectral function broadening by the Gaussian broadening
kernel. The agreement of the calculated spectral function within the
conduction band, i.e. in the $[-D:D]$ interval, Fig.~\ref{fig55}b, with the
exact result is also very satisfactory.

\section{High-resolution spectral functions} \label{secexample}

We present two examples of high-resolution calculations unmasking
interesting details. We first study the emergence of the Kondo resonance as
the electron-electron repulsion $U$ is increased in the Anderson impurity
model. We then consider the Anderson-Holstein model to show that the
phononic side peaks can be well resolved.

\subsection{The emergence and the shape of the Kondo resonance}

\begin{figure}[htbp]
\includegraphics[width=8cm,clip]{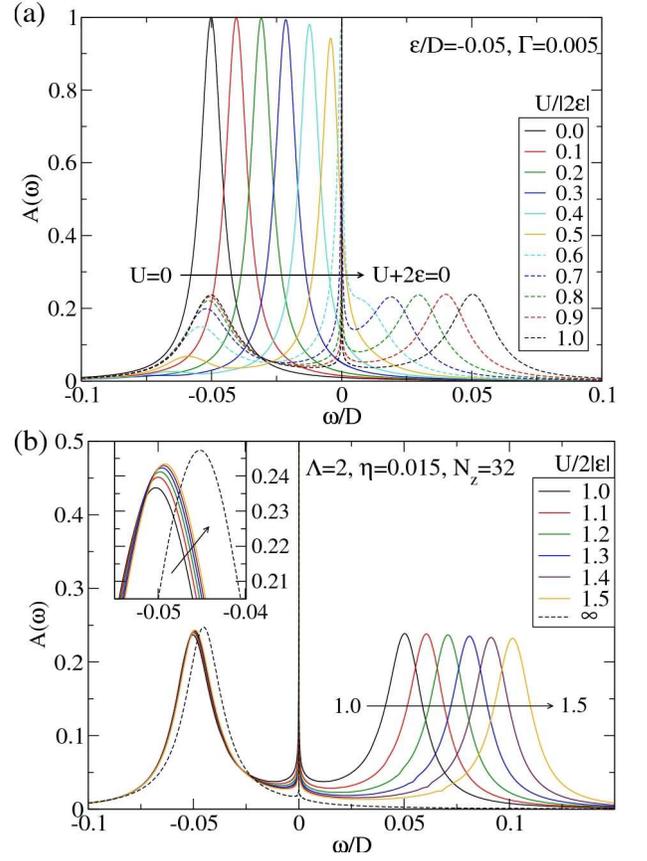}
\caption{(Color online) Spectral functions of the Anderson model for
increasing $U$.}
\label{fig2}
\end{figure}

In Fig.~\ref{fig2} we show spectral functions for the Anderson impurity
model for a range of values of the electron-electron interaction $U$, from
the non-interacting case to the symmetric situation $U=-2\epsilon$
(Fig.~\ref{fig2}a) and then to the large-$U$ limit (Fig.~\ref{fig2}b). Since
these results are hardly affected by overbroadening, we can accurately
follow the evolution of the spectral peak, its location as well as its
height and width. In the non-interacting limit, the peak height is
$1/\pi\Gamma$, its width is $\approx \Gamma$ and it is centered at $\omega
\approx \epsilon$. As $U$ increases, the peak position shifts linearly with
$U$ (Hartree shift), while its height decreases. The remaining spectral
weight is located in the emerging lower charge-transfer spectral peak (i.e.
the lower ``Hubbard band''); this peak is initially located below
$\epsilon$, but it shifts to $\approx \epsilon$ as we approach the
particle-hole symmetric situation. The width of the charge-transfer peaks is
roughly twice ($2\Gamma$) the width of the original non-interacting peak
($\Gamma$). As we increase $U$ further, Fig.~\ref{fig2}b, the lower
charge-transfer peak shifts only weakly as a function of $U$, while the
upper charge-transfer peak shifts as $\epsilon+U$; in the range of finite
$U$ shown, its height decreases only slightly and the width remains nearly
constant. At the same time, the width of the Kondo resonance is
significantly reduced, but we find that it remains almost pinned at the
Fermi level (at $U=\infty$, for example, the half-width at half-maximum of
the Kondo peak is $1.2\, 10^{-8}D$, while the shift of the maximum is only
$3.6\, 10^{-10}D$, i.e. 3 percent in the units of HWHM). This is in
agreement with the Fermi liquid theory, but in disagreement with the results
from methods based on the large-$N$ expansion, such as the non-crossing
approximation, which overestimate the shift of the resonance, in particular
for $N=2$. It also implies that the Kondo temperature should better not be
defined as the displacement of the Kondo resonance from the Fermi level, as
it is sometimes done.

\begin{figure}[htbp]
\includegraphics[width=8cm,clip]{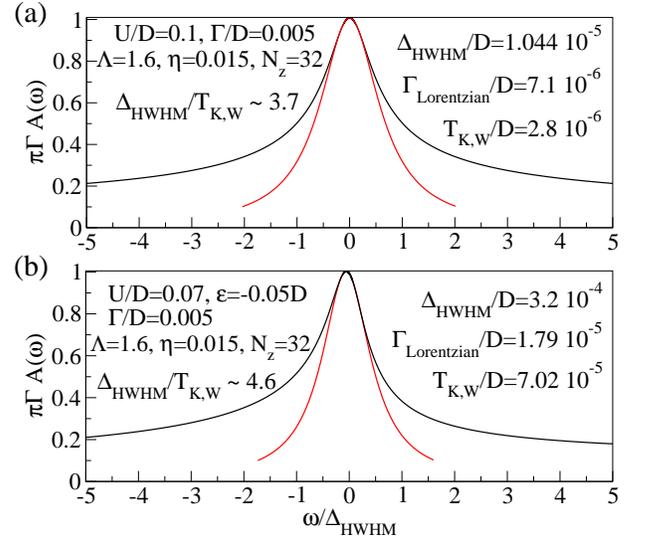}
\caption{(Color online) Close-up on the Kondo resonance of a) symmetric and
b) asymmetric Anderson impurity model and a fit to a Lorentzian (red curve)
in the Fermi liquid regime for $\omega \ll T_K$.} \label{fig3}
\end{figure}

In Fig.~\ref{fig3}a we plot a close-up on the Kondo resonance in the
symmetric case, $\epsilon+U/2=0$. As expected, the peak shape deviates
significantly from a Lorentzian shape \cite{bulla2000lm, dickens2001,
glossop2002, rosch2003}. In fact, true agreement is only found in the
asymptotic $\omega \to 0$ region, where both the Lorentzian curve and the
spectral function have quadratic frequency dependence. In the latter case,
this is mandated by the Fermi-liquid behavior at low energy scales.

The relation between the width of the Kondo resonance and the Kondo
temperature $T_K$ (times $k_B$) is of considerable experimental interest, in
particular for tunneling spectroscopy. In the symmetric case, we find for
the ratio between the half-width at half-maximum and the Kondo temperature
(Wilson's definition):
\begin{equation}
\Delta_\mathrm{HWHM}/T_{K,W} \approx 3.7.
\end{equation}
The Kondo temperature $T_{K,W}$ is defined as $\chi_\mathrm{imp}(T=0) =
(g\mu_B)^2 (W/4\pi) 1/k_B T_{K,W} \approx (g\mu_B)^2 0.103/k_BT_{K,W}$, where
$W=e^{C+1/4}/\sqrt{\pi}$ is Wilson number, and extracted from the NRG
results for the magnetic susceptibility using the prescription $k_B T_{K,W}
\chi_\mathrm{imp} (T_{K,W})/(g\mu_B)^2 = 0.07$. The same value of 3.7 is
also obtained when log-Gaussian broadening is used with a small value of the
parameter $b$ and with suitable $z$-averaging. This ratio is lower than some
other values reported in the literature \cite{micklitz2006}.

In Fig.~\ref{fig3}b we plot the Kondo resonance in the asymmetric case. We
find that the ratio between the half-width and the Kondo temperature is now
$\Delta_\mathrm{HWHM}/T_{K,W}=4.6$. Even though we are still deep in the
Kondo regime (phase shift is $\delta \approx 0.47 \pi$), the Kondo peak has
developed a significant asymmetry in its shape. These line-shape effects are
important in the interpretation of experimental results. Due to
uncertainties in the ratio $\Delta_\mathrm{HWHM}/T_{K,W}$, the expected
systematic error in determining $T_K$ from the Kondo peak width is estimated
to be several 10 percents. This implies that comparisons of $T_K$ of
different adsorbate/surface systems determined in this way are rather
meaningless unless the differences are of the order of a factor 2 or more.

\subsection{The phononic side-peaks in the Anderson-Holstein model}

We consider the Anderson-Holstein model with coupling of a local Einstein
phonon mode to charge fluctuations:
\begin{equation}
H_\mathrm{imp} = \epsilon\, n+U n_\uparrow n_\downarrow
+g(n-1)(a^\dag+a) + \omega_0 a^\dag a.
\end{equation}
Here $a$ is the bosonic phonon operator, $\omega_0$ is the oscillator
frequency and $g$ the coupling between the impurity charge and the
oscillator displacement. This model was studied intensely using a variety of
techniques, including NRG \cite{hewson2002, jeon2003, cornaglia2004,
cornaglia2005, spincharge}. Its applications range from the problem of small
polaron and bipolaron formation, electron-phonon coupling in heavy fermions
and valence fluctuation systems, to describing the electron transport
through deformable molecules.

The effect of the electron-phonon coupling is to reduce the effective
electron-electron interaction and shift the level energy \cite{mahan,
hewson2002}:
\begin{equation}
\begin{split}
U_\mathrm{eff} &= U-2\frac{g^2}{\omega_0}, \\
\epsilon_\mathrm{eff} &= \epsilon+\frac{g^2}{\omega_0}.
\end{split}
\end{equation}
In addition, the effective hybridisation becomes phonon-dependent, since the
phonon cloud can be created or absorbed when the impurity occupancy changes
\cite{hewson2002}.

\begin{figure}[htbp]
\includegraphics[width=8cm,clip]{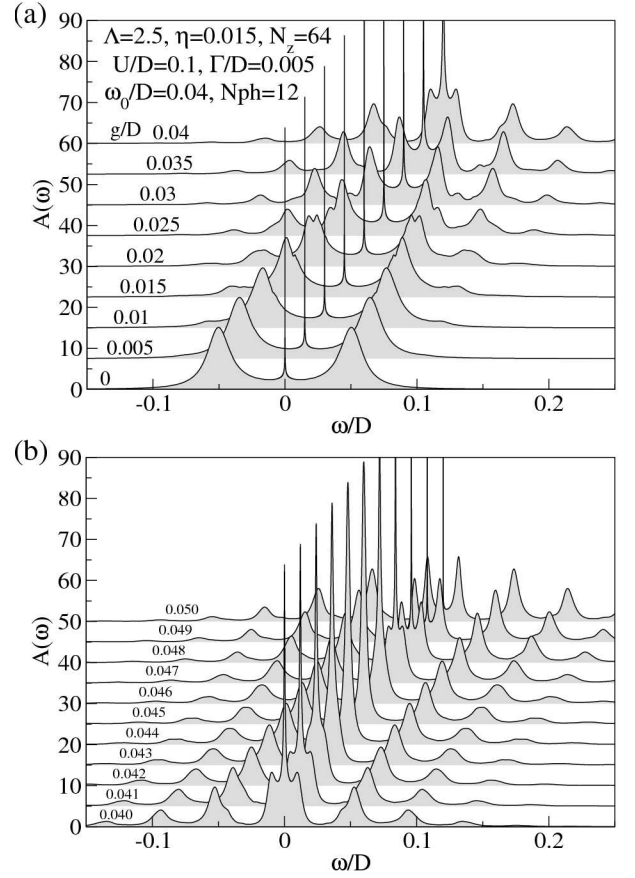}
\caption{Spectral functions for the Anderson-Holstein model in the
particle-hole symmetric case, $\delta=\epsilon+U/2=0$.} 
\label{fig10}
\end{figure}

It is possible to resolve the phononic side-peaks and the transition to the
charge Kondo regime, Fig.~\ref{fig10}. For small coupling $g$, we see the
gradual emergence of the phononic side-peaks at energies
$\epsilon_\mathrm{eff} + U_\mathrm{eff} + n\omega_0$, $n=1,2,3,\ldots$. In
addition to these peaks, we see that the charge transfer peak at
$\epsilon_\mathrm{eff}+U_\mathrm{eff}$ itself has internal structure; as $g$
increases, part of the spectral weight is transferred from this peak to
higher energies in the form of a smaller peak which eventually merges with
the first phononic side-peak at
$\epsilon_\mathrm{eff}+U_\mathrm{eff}+\omega_0$. The transition from spin to
charge Kondo regime occurs at $g/D \approx 0.0445$, when $U_\mathrm{eff}
\approx 0$. At the transition, the charge transfer peak merges with what
used to be the Kondo resonance to give a single broad resonance whose width
is no longer set by the energy scale of the Kondo effect, but rather by some
renormalized spectral width $\Gamma_\mathrm{eff}$.

\section{NRG as a high-resolution impurity solver for DMFT}
\label{secdmft}

\begin{figure}[htbp]
\includegraphics[width=8cm,clip]{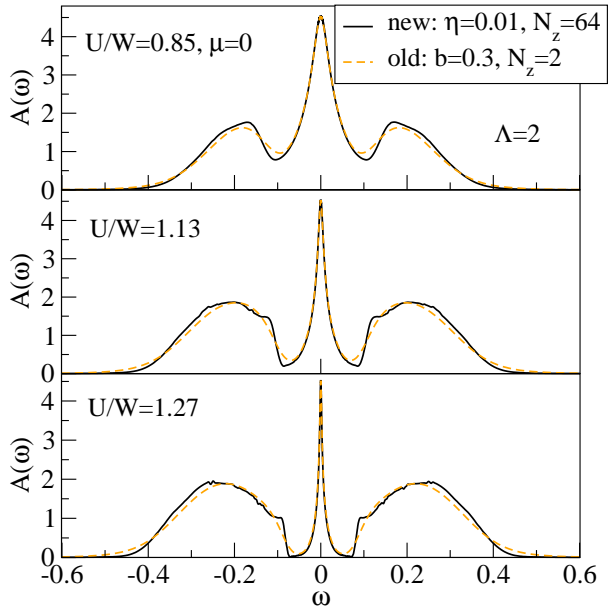}
\caption{Local spectral functions of the Hubbard model on the hypercubic lattice.}
\label{hyper}
\end{figure}

The most severe shortcoming of the NRG (using log-Gaussian broadening with
large $b$ and traditional discretization schemes) in its role as an impurity
solver in DMFT was the reduced energy resolution at finite excitation
energies. This not only affects the self-consistent calculation by
introducing systematic errors, but sometimes features in spectral functions
at high energies (for example kinks in the excitation dispersions) are
themselves of interest. We demonstrate the applicability of the new approach
on the simplest example of the Hubbard model. The case of hypercubic lattice
is considered in Fig.~\ref{hyper} where we plot the local density of states
for a range of the repulsion parameter $U$ as the metal-insulator transition
is approached. Compared to the results computed using the conventional NRG
approach, the high-energy features (Hubbard bands) are sharper. Furthermore,
the conventional approach underestimates the reduction of the density of
states (``pseudo-gap'') between the Hubbard bands and the quasiparticle
peak. We also observe that the Hubbard bands have inner structure. We find a
notable peak at the inner edges of the Hubbard band; the existence of some
spectral features at the band edges had been suggested already in the early
iterative perturbation theory, non-crossing approximation, quantum Monte
Carlo and NRG DMFT results for the Hubbard model and the existence of a
sharp peak was demonstrated in more recent high-resolution dynamic
density-matrix renormalization (D-DMRG) calculations \cite{karski2005,
karski2008}. 

\begin{figure}[htbp]
\includegraphics[width=8cm,clip]{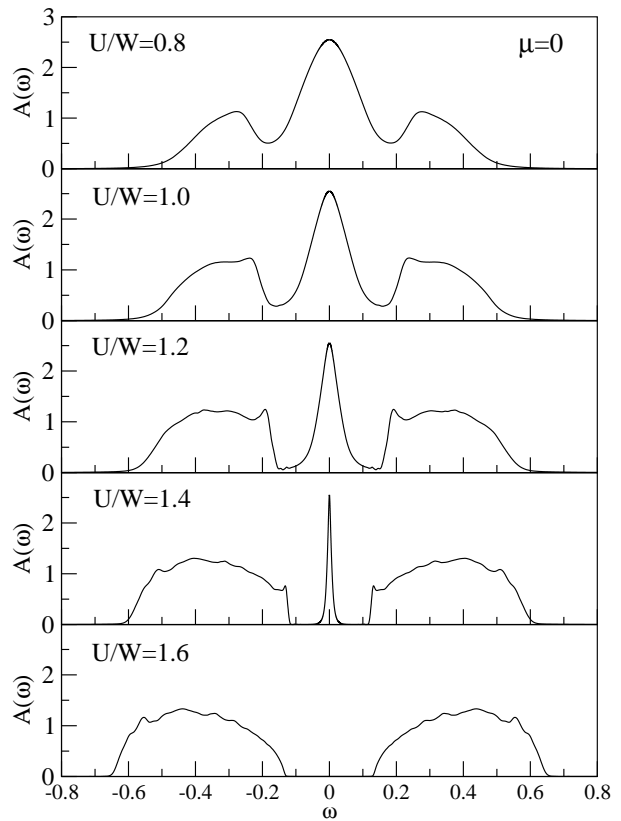}
\caption{Local spectral functions of the Hubbard model on the Bethe lattice.}
\label{bethe}
\end{figure}

On the Bethe lattice, Fig.~\ref{bethe}, the Hubbard bands are sharper due to
the finite support of the lattice density of states and the inner Hubbard
band edge peaks are sharper. There are furthermore less pronounced spectral
features at integer multiples of the energy of the inner Hubbard band edge;
they are most visible in the $U/W=1.4$ results. We also calculated the local
two-particle Green functions at the end of the DMFT cycle to try to obtain
some insight whether these additional structures are possibly related to
certain two-particle excitations. However, no clear evidence was found for
such statement. Thus, at present, we cannot give a satisfactory physical
explanation of these additional structures. In any case, they motivate
further high-resolution studies of both the single-impurity Anderson model
and the Hubbard model in DMFT, concentrating on the regime with vanishing
Kondo resonance.

\section{Conclusion}

We presented spectral function calculations which indicate that the
numerical renormalization group method allows to compute more accurate
results than it is generally believed. We have shown that overbroadening
effects can be in large part removed by using a sufficiently narrow Gaussian
broadening kernel. Furthermore, we have shown that there is surprisingly
little variation as $\Lambda$ is decreased (disregarding the artefact
shifts), thus there is no inherent overbroadening due to the discretization
of the conduction band. At best one can say that as $\Lambda$ is increased,
more $z$ values need to be used in the interleaved method to obtain smooth
spectral functions. It must be emphasized that sweeping over $z$ is an
embarrassingly parallel problem, i.e. essentially no overhead is associated
with splitting the problem into a large number of parallel tasks.

As the continuum limit is approached ($\Lambda \to 1$), the discretization
artefacts in the spectral function calculated using the traditional schemes
shift out toward the band-edge, but in the range of $\Lambda$ that can be
used in practical calculations, the artefacts are always present. The use of
the logarithmic discretization is commonly justified by the rapid
convergence of calculated quantities to the continuum limit; while static
properties indeed converge rapidly, this is not the case with dynamic
properties. The presence of artefacts therefore has several implications for
NRG calculation. First of all, in the traditional approach it cannot be
claimed that a calculation is performed for a given density of states
$\rho(\omega)$, but rather for a band with a density of states given by
$A_{f_0}(\omega)$ in the problem with decoupled impurity. The presence of
the structure in the spectral function $A_{f_0}(\omega)$ then forcibly leads
to what is perceived as ``artefacts'' in the impurity spectral function
$A(\omega)$. Artefacts have important implications for the application of
the NRG in DMFT, since these anomalies lead to features in the impurity
spectral function that are difficult to disassociate from real fine
structure. A good test to distinguish between artefacts and real spectral
features is to perform calculations for several values of $\Lambda$, keeping
all other parameters constant. Real features will change very little, while
artefacts will shift and change form significantly. Depending on the
circumstances (structure of the impurity model, model parameters) and the
purposes (single-impurity calculation vs. self-consistent dynamical
mean-field-theory calculation), the artefacts are either benign or rather
detrimental. 

The proposed new way of calculating the coefficients $\Epsilon_j^z$ leads to
a sizable improvement in the convergence to the $\Lambda \to 1$ limit and
to a significant reduction of the discretization artefacts. Since the DMFT
self-consistency loop couples low-energy and high-energy scales, the 
reduction of the artefacts at high energies is a significant improvement
which increases the reliability of the NRG as an impurity solver.

\begin{acknowledgments}
We thank Janez Bon\v{c}a for providing the motivation which led to this work
and Robert Peters for discussions on the calculation of spectral functions.
We acknowledge computer support by the Gesellschaft f\"ur wissenschaftliche
Datenverarbeitung (GWDG) in G\"ottingen and support by the German Science
Foundation through SFB 602.
\end{acknowledgments}

\appendix

\section{Spectral function broadening}
\label{app0}

\begin{figure}[ht!]
\includegraphics[width=8cm,clip]{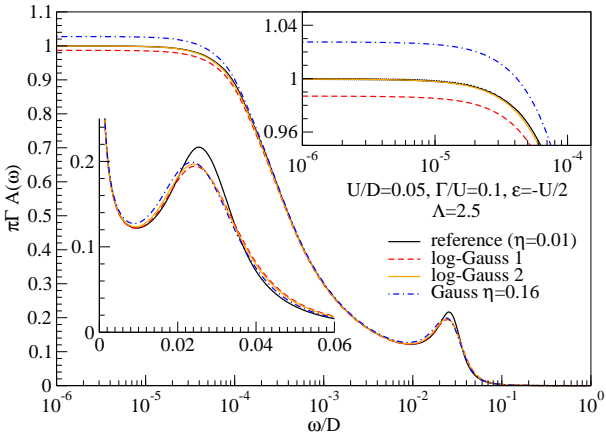}
\caption{Spectral function of the symmetric Anderson impurity model:
comparison of results obtained using different broadening kernels. Reference
results are calculated using Gaussian broadening with sufficiently narrow
kernel so that very little change is obtained by further narrowing.}
\label{fig9}
\end{figure}

In practical NRG calculations, $z$-averaging is performed over a smaller
number of twist parameters, therefore wider broadening functions must be
used. It is thus interesting to compare the results for spectral functions
obtained using different broadening kernels, Fig.~\ref{fig9}. We compare
simple Gaussian broadening, conventional log-Gaussian broadening and a
modified log-Gaussian kernel proposed in Ref.~\onlinecite{weichselbaum2007}:
\begin{equation}
\label{alpha}
P(\omega,E) = \frac{1}{\sqrt{\pi} \alpha |E|}
e^{-\left[ \frac{\ln \omega-\ln E}{\alpha} - \gamma \right]^2}
\end{equation}
with $\gamma=\alpha/4$. The same $\Lambda$, $p$ and $N_z$ were use for all
three broadening kernels, with $b=\alpha=\sqrt{2}\eta$. 

In the low-energy ($\omega \ll T_K$) range, we find that Gaussian broadening
overestimates the spectral density, log-Gaussian broadening underestimates
it, while the modified log-Gaussian kernel, Eq.~\eqref{alpha}, gives a very
good approximation to the high-resolution result. All three approaches
describe quite well the flanks of the Kondo resonance. Gaussian broadening
overestimates the spectral density in the energy range between the Kondo
resonance and the Hubbard peak, while the best results are here obtained by
the original log-Gaussian broadening. All three broadening approaches shift
the maximum of the Hubbard peak to lower energies to roughly comparable
degree. Finally, in the high-energy range, log-Gaussian approaches
overestimate the spectral density more than the simple Gaussian broadening.

For studying low-energy properties with typical NRG broadening parameters,
the modified log-Gaussian kernel, Eq.~\eqref{alpha}, is the best choice. For
high-resolution studies with very small broadening, all three broadening
techniques become almost equivalent, but the plain Gaussian kernel has a
small advantage by being symmetric; the symmetry leads to smaller deviations
of higher-moment spectral sum rules.

\section{Modified discretization scheme}
\label{app1}

We describe the modified discretization scheme which consists
of solving the ordinary differential equation for $\Epsilon_j^z$:
\begin{equation}
\label{eqa1}
\frac{\int_{I_j} \rho(\epsilon) d\epsilon}
{|d\Epsilon_j^z/dz|} = \rho(\omega).
\end{equation}
As a first step, we introduce continuous indexing as $x=j+z$ with parameter
$x$ running from 1 to $+\infty$, so that coefficients $\Epsilon_j^z$ and
$\epsilon_j^z$ become continuous functions of $x$, i.e. $\Epsilon(x)$ and
$\epsilon(x)$. We then rewrite Eq.~\eqref{eqa1} as
\begin{equation}
\frac{d\Epsilon(x)}{dx}= \frac{\int_{\epsilon(x)}^{\epsilon(x+1)}
\rho(\omega)d\omega }{\rho[\Epsilon(x)]}
\end{equation}
with the initial condition $\Epsilon(1)=D$. It is helpful to take into
account the expected asymptotic behavior of $\Epsilon(x)$ using
the following Ansatz:
\begin{equation}
\Epsilon(x)=D f(x) \Lambda^{2-x},
\end{equation}
with $f(1)=1/\Lambda$. The equation to solve is then
\begin{equation}
\frac{df(x)}{dx} = \ln\Lambda \, f(x) - 
\frac{\int_{\epsilon(x+1)}^{\epsilon(x)} \rho(\omega)d\omega}
{\Lambda^{2-x} \rho[\Epsilon(x)]}.
\end{equation}
This equation can be solved numerically; for general $\rho(\omega)$ it is
advisable to use arbitrary-precision numerics for this purpose, since the
equation is stiff. For DOS which is finite at the Fermi level, we must have
$f(\infty) = (1-\Lambda^{-1})/\ln\Lambda$. Checking the convergence to this value
is a good test of the integration procedure.

\bibliography{paper}

\end{document}